\documentclass[11pt,a4paper]{article}
\usepackage{amsmath,amssymb,color}
\usepackage{epsfig,graphicx}
\usepackage{scalefnt}
\usepackage{mciteplus}
\mciteSetMidEndSepPunct{;\\}{.}{\relax}
\usepackage{setspace}
\usepackage{hyperref}

\begin{document}

\begin{center}
\hspace*{\fill}{DESY-17-152}\\
\hspace*{\fill}{FR-PHENO-2017-018}\\
\hspace*{\fill}{IPPP/17/72}\\
\hspace*{\fill}{TTP17-039}\\
\vspace*{1cm}
  {\Large\bf Addendum to ``Charm and Bottom Quark Masses: An Update''}

  \vspace*{3em}

  Konstantin G. Chetyrkin$^a$,
  Johann H. K\"uhn$^b$,
  Andreas Maier$^c$,
  Philipp Maierh\"ofer$^d$,
  \\
  Peter Marquard$^e$,
  Matthias Steinhauser$^b$
  and Christian Sturm$^f$
  \\[.5em]
  $(a)$ \small{\em II. Institute for Theoretical Physics, Hamburg University,}\\
  {\small\em 22761 Hamburg, Germany}
  \\[.3em]
  $(b)$ \small{\em Institut f\"ur Theoretische Teilchenphysik,}\\
  {\small\em Karlsruhe~Institute~of~Technology~(KIT), 76128 Karlsruhe, Germany}
  \\[.3em]
  $(c)$ \small{\em IPPP, Department of Physics, University of Durham,}\\
  {\small\em DH1 3LE, United Kingdom}
  \\[.3em]
  $(d)$ \small{\em Physikalisches Institut, Albert-Ludwigs-Universit\"at Freiburg,}\\
  {\small\em 79104 Freiburg, Germany}
  \\[.3em]
  $(e)$ \small{\em Deutsches Elektronen-Synchrotron, DESY,}\\
  {\small\em 15738 Zeuthen, Germany}
  \\[.3em]
  $(f)$ 
  \small{\em Institut f\"ur Theoretische Physik und Astrophysik,}
  {\small\em Universit\"at W\"urzburg, Emil-Hilb-Weg 22, D-97074 W\"urzburg, Germany}

  \vspace{1em}

  \begin{minipage}{30em}
    We update the experimental moments for the charm quark as computed in
    Ref.~\cite{Kuhn:2007vp} and used in Refs.~\cite{Chetyrkin:2009fv}
    and~\cite{Chetyrkin:2010ic} for the determination of the charm-quark
    mass. The new value for the $\overline{\rm MS}$ charm-quark mass reads
    $m_c(3\, \text{GeV})=0.993\pm0.008$~GeV.
  \end{minipage}

  \vspace{1em}

\end{center}

In Ref.~\cite{Chetyrkin:2009fv} the $\overline{\rm MS}$ charm- and
bottom-quark masses have been determined using relativistic sum
  rules which relate theoretically calculated moments of the photon vacuum
polarization function to experimentally measured moments. The latter are
determined from measurements of the $R$-ratio and properties of the narrow
resonances. The moments of the vacuum polarization function can be computed in
perturbative QCD.  In this note we update the experimental input and
re-evaluate the corresponding moments. New results for the charm-quark mass
$m_c$ are presented which are about 35\% more precise than those of
our previous determination.

For convenience we briefly present the formalism which is used
in order to obtain $m_c$.  The $n$-th theory moment
is obtained from
\begin{equation}
  {\cal M}_n^{\rm th} = \frac{12\pi^2}{n!}
  \left(\frac{{\rm d}}{{\rm d}q^2}\right)^n
  \Pi_c(q^2)\Bigg|_{q^2=0}
  \,,
\label{eq:Mtheo}
\end{equation}
where $\Pi_c(q^2)$ is the vector-current correlator with virtual
charm-quark loops which can be cast into the form
\begin{equation}
  \Pi_c(q^2) = Q_c^2 \frac{3}{16\pi^2} \sum_{n\ge0}
                       \bar{C}_n z^n
  \,,
  \label{eq:pimom}
\end{equation}
with $z=q^2/(4m_c^2)$. Here $m_c=m_c(\mu)$ is the $\overline{\rm
  MS}$ heavy quark mass at the scale $\mu$ and $Q_c=2/3$
is the electric charge of the charm quark in units of the
elementary charge. The results which we use for the first four
coefficients  $\bar{C}_n$ are known up to four-loop accuracy
analytically~\cite{Chetyrkin:2006xg,*Boughezal:2006px,*Maier:2008he,*Maier:2009fz,*Maier:2017xx}. For applications and calculational techniques of the
determination of the related massive tadpole diagrams we refer to the
review~\cite{Chetyrkin:2015mxa}.
Equating the theory moments ${\cal M}_n^{\rm th}$ with the
experimentally measured moments,
\begin{equation}
  \label{eq:M_exp}
  {\cal M}_n^{\text{exp}}=\int \frac{{\rm d}s}{s^{n+1}}R_c(s)
  \,,
\end{equation}
where $R_c=\sigma(e^+e^-\to c\bar{c})/\sigma(e^+e^-\to \mu^+\mu^-)$,
leads to
\begin{equation}
  \label{eq:m_Q}
  m_c= 
  \frac{1}{2}\left(
    \frac{9Q_c^2}{4}\frac{\bar{C}_n}{{\cal M}_n^{\text{exp}}}
  \right)^{\frac{1}{2n}}\,,
\end{equation}
which can be used in order to extract the
charm-quark mass. The experimental moments ${\cal
  M}_n^{\text{exp}}$ receive contributions from the narrow resonances,
the charm-threshold region and the continuum region above
a center of mass energy $\sqrt{s}$ of about $5$~GeV.  Even for small
values of $n$ the contributions from the $J/\Psi$ and $\Psi(2S)$
resonances are dominant.

There is essential new input from measurements of the electronic
decay width $\Gamma_{ee}$ of the
$J/\Psi$~\cite{Ablikim:2016xbg} and $\Psi(2S)$~\cite{Ablikim:2015ain}
resonances from BES~III which shall be used in the
following. The latter value is incorporated into the latest PDG
result~\cite{Patrignani:2016xqp}, whereas $\Gamma_{ee}(J/\Psi)$
of Ref.~\cite{Ablikim:2016xbg} is not included. We thus
combine the results from Refs.~\cite{Ablikim:2016xbg}
and~\cite{Patrignani:2016xqp} and obtain the updated resonance input
parameters as listed in Tab.~\ref{tab::gamma}.  We also update the mass
values for the resonances using the recent PDG
values~\cite{Patrignani:2016xqp}. Note, however, that their
improvement has no influence on the results for the moments.
Furthermore we update the value of the strong coupling constant
and use $\alpha_s(M_Z)=0.1181\pm0.0011$~\cite{Patrignani:2016xqp}
(instead of $\alpha_s(M_Z)=0.1189\pm0.002$ as in
Ref.~\cite{Chetyrkin:2009fv}).

\begin{table}[t]
  \begin{center}
    \begin{tabular}{c|c|c}
      \hline
      &$J/\Psi$     &$\Psi(2S)$    \\ \hline
      $M_{\Psi}$(GeV)~\cite{Patrignani:2016xqp}    & 3.096900(6) & 3.686097(25)  \\
      $\Gamma_{ee}$(keV)~\cite{Ablikim:2016xbg,Patrignani:2016xqp} & 5.57(8)  & 2.34(4)      \\
      $(\alpha/\alpha(M_{\Psi}))^2$& 0.957785     & 0.95554      \\
      \hline
    \end{tabular}
    \caption{\label{tab::gamma}Updated input values for the resonance parameters.}
  \end{center}
\end{table}

For the moments we obtain

\begin{center}
\begin{tabular}{l|lll|l||ll}
\hline
$n$ & ${\cal M}_n^{\rm res}$
& ${\cal M}_n^{\rm cc}$
& ${\cal M}_n^{\rm cont}$
& ${\cal M}_n^{\rm exp}$
& ${\cal M}_n^{\rm np}$
\\
\hline
 & $\times 10^{(n-1)}$
& $\times 10^{(n-1)}$
& $\times 10^{(n-1)}$
& $\times 10^{(n-1)}$
& $\times 10^{(n-1)}$
\\
\hline
$1$&$  0.1191(14)$ &$  0.0318(15)$ &$  0.0645(10)$ &$  0.2154(23)$ &$ -0.0001(3)$ \\
$2$&$  0.1169(15)$ &$  0.0178(8)$ &$  0.0143(3)$ &$  0.1490(17)$ &$ -0.0002(5)$ \\
$3$&$  0.1165(15)$ &$  0.0101(5)$ &$  0.0042(1)$ &$  0.1308(16)$ &$ -0.0004(8)$ \\
$4$&$  0.1176(16)$ &$  0.0058(3)$ &$  0.0014(0)$ &$  0.1248(16)$ &$ -0.0006(12)$ \\
\hline
\end{tabular}
\end{center}

and the updated table for the charm-quark mass reads

\begin{center}
\begin{tabular}{l|c|llll|l|l}
\hline
$n$ & $m_c(3~\mbox{GeV}) $ & 
exp & $\alpha_s$ & $\mu$ & ${\rm np}_{\rm LO}$& 
total   & $m_c(m_c)$
\\
\hline
         1&  0.993&  0.007&  0.004&  0.002&  0.001&  0.008&  1.279 \\
         2&  0.982&  0.004&  0.007&  0.005&  0.001&  0.010&  1.269 \\
         3&  0.982&  0.003&  0.008&  0.006&  0.001&  0.010&  1.269 \\
         4&  1.003&  0.002&  0.005&  0.028&  0.001&  0.029&  1.288 \\
\hline
\end{tabular}
\end{center}

where ${\rm np}_{\rm LO}$ indicates that we use the leading order (LO)
approximation for the gluon condensate contribution (see also the
discussion in Ref.~\cite{Chetyrkin:2009fv}).

One observes a noteworthy reduction of the uncertainty in the
experimental moments.  As compared to the results from
Ref.~\cite{Chetyrkin:2009fv} there is an increase in the
charm-quark mass by 7~MeV for $n=1$, by 6~MeV for $n=2$, by
4~MeV for $n=3$ and a decrease by $1$~MeV for $n=4$. For
$n=1$, which constitutes our final result, the uncertainty decreases
from $13$~MeV to $8$~MeV.  Within the uncertainty all results in the
above table are consistent with each other and with the results obtained
in Ref.~\cite{Chetyrkin:2009fv}.  Our final result for the
$\overline{\rm MS}$ charm-quark mass reads $m_c(3\,
\text{GeV})=0.993\pm0.008$~GeV and
$m_c(m_c)=1.279\pm0.008$~GeV.

\section*{Acknowledgments}
We would like to thank Achim Denig for clarifications concerning
Refs.~\cite{Ablikim:2016xbg} and~\cite{Ablikim:2015ain} and Ruth Van De Water
for questions initiating this new analysis. This work is supported by the BMBF
through Grant No. 05H15VKCCA. 

\begin{spacing}{0.5}
\providecommand{\href}[2]{#2}\begingroup\raggedright\endgroup
\end{spacing}

\end{document}